\documentclass[12pt]{article}
\usepackage{amssymb,amsmath,epsfig}
\usepackage{graphicx}
\allowdisplaybreaks
\begin{document}
\title{\textbf{Comprehensive Study of Generalized Ghost Dark Energy
in $f(\textsl{Q}, \textsl{L}_{m})$ Gravity: New Insights into Cosmic
Dynamics}}
\author{M. Zeeshan
Gul$^{1,2}$\thanks{mzeeshangul.math@gmail.com,
zeeshan.gul@khazar.org}, M.
Sharif$^1$\thanks{msharif.math@pu.edu.pk}, Shan Ali Qureshi
$^1$\thanks{shanali.math@gmail.com} and Baiju Dayanandan$^3$ \thanks{baiju@unizwa.edu.om}\\
$^1$ Department of Mathematics and Statistics, The University of Lahore,\\
1-KM Defence Road Lahore-54000, Pakistan.\\
$^2$ Research Center of Astrophysics and Cosmology, Khazar
University,\\ Baku, AZ1096, 41 Mehseti
Street, Azerbaijan.\\$^3$Natural and Medical Sciences Research
Center,\\ University of Nizwa, Oman.}

\date{}
\maketitle

\begin{abstract}
This paper explores the generalized ghost dark energy model in the
framework of $f(\textsl{Q}, \textsl{L}_{m})$ gravity, where
$\textsl{Q}$ represents the non-metricity scalar and
$\textsl{L}_{m}$ denotes the matter-Lagrangian density. We take the
homogeneous and isotropic universe with an ideal matter distribution
and examine a scenario with interacting dark energy and dark matter.
We then reconstruct $f(\textsl{Q}, \textsl{L}_{m})$ model to examine
the effects of this extended gravitational framework on the cosmic
evolution. The behavior of numerous cosmic parameters are explored
corresponding to distinct parametric values. The stability is
evaluated by the squared sound speed method. The statefinder $(r,s)$
and standard diagnostic pairs $(\omega_D-\omega'_{D})$ are used to
study the various cosmic eras. Our results align with recent
observational evidence, indicating that the $f(\textsl{Q},
\textsl{L}_{m})$ model effectively characterizes dark energy and
cosmic evolution.
\end{abstract}

\section{Introduction}

The mysterious nature of the cosmos and astrophysical events has
captivated numerous scholars to explore and study these phenomena.
The cosmos consists of three major components, i.e., dark energy
(\textsl{DE}), dark matter (\textsl{DM}) and normal matter. The
universe is primarily governed by \textsl{DE} and \textsl{DM} with
usual matter occupies the remaining space. The \textsl{DM} is an
invisible matter and its existence is deducted by gravitational
lensing and the rotation curves of galaxies \cite{1}. However, the
greatest ground-breaking finding of recent decades has been the
accelerated expansion of the cosmos, which has steered scientific
inquiry in a completely new direction. A mysterious source of energy
which exhibits significant negative pressure, referred to as
\textsl{DE} is thought to drive this expansion. The challenges
surrounding the properties and existence of both \textsl{DE} and
\textsl{DM} remain some of the complex and unsolved challenges in
cosmology. The significant model for elucidating the characteristics
of \textsl{DE} is the $\Lambda \textsl{CDM}$ framework. In this
model, the cosmological constant is regarded as the most probable
explanation for the accelerating expansion of the universe. Although
it aligns well with observational data, but it encounters challenges
such as fine-tuning and the cosmic coincidence problem \cite{2}.
However, there are two methodologies to comprehend enigmatic
features of the cosmos. One approach involves modifying the
geometric aspect of the Einstein-Hilbert action, resulting in
alternative theories of gravity, while the other focuses on altering
the matter component to create dynamical \textsl{DE} models
\cite{3}.

Researchers have proposed various methods to address these
challenges in the past decades, yet they are still mysterious. To
understand the nature of \textsl{DE}, the Veneziano ghost dark
energy (\textsl{GDE}) model has proposed in \cite{4} which has
notable physical effects. It generates a tiny vacuum energy density
in curved spacetime corresponding to $\Lambda^{3}_{\textsl{QCD}}$,
where \textsl{QCD} represents the quantum chromodynamics. Therefore,
no additional parameters, degrees of freedom or alterations are
needed. With $\Lambda_{\textsl{QCD}} \sim 100MeV$ and $H \sim
10^{-33}eV$, $\Lambda^{3}_{\textsl{QCD}}$ provides the proper order
for the \textsl{DE} density. This shows that this paradigm
eliminates the fine-tuning problem \cite{5}. Cai et al \cite{6}
investigated the cosmic acceleration corresponding to \textsl{DE}
model. Sheykhi and Movahed \cite{7} examined cosmic evolution by
analyzing the \textsl{GDE} model. The universe rapid growth via the
\textsl{GDE} model has been investigated in \cite{8}.

Einstein's gravitational theory (\textsl{GR}) has been modified into
various alternative gravitational theories due to their significant
interest in explaining the accelerated expansion of the universe.
The equivalent geometric frameworks can be used to represent
\textsl{GR}. The initial approach focuses on curvature with the
absence of both torsion and non-metricity. The second is the
teleparallel formalism, where both non-metricity and curvature are
absent. An alternative representation is also possible in which
gravitational effects are characterized through the non-metricity of
the metric, which reflects variations in a vector's length as it
undergoes parallel transport. The fundamental principle of
teleparallel gravity is to substitute the spacetime metric with a
set of tetrad vectors, which introduces torsion. Additionally,
curvature is substituted with the torsion produced by the vierbein,
which serves to characterize the gravitational influences in the
cosmos \cite{13}. Linder \cite{14} introduced modified teleparallel
gravity ($f(\mathbb{T})$), where $\mathbb{T}$ denotes the torsion
scalar. Jimenez et al \cite{17} introduced the concept of symmetric
teleparallel gravity, known as $f(\textsl{Q})$ gravity. A lot of
significant work has been done in this modified framework
\cite{17a}-\cite{17g}.

Another modified proposal is $f(\textsl{Q},\textsl{T})$ theory,
which has become subject of great interest in scientific community
due to its crucial implications in the field of cosmology and
astrophysics \cite{18a}-\cite{18k}. Alternative theories and
observational constraints has been examined in \cite{19a}-\cite{19g}
Myrzakulov et al \cite{23} generalized the symmetric teleparallel
theory by including the matter-Lagrangian in the action, known as
$f(\textsl{Q}, \textsl{L}_{m})$ theory. An important aspect of this
gravity is that the corresponding field equations are of
second-order which make this theory distinct from $f(\textsl{R})$
gravity ($\textsl{R}$ represents the curvature invariant), which
employs fourth-order field equations. This modified proposal
investigates its potential consequences, its consistency with
present experimental evidence and determine its applicability to
cosmological models. The $f(\textsl{Q}, \textsl{L}_{m})$ theory
establishes a unique relationship between geometry and matter,
garnering a profound interest because of its significant
consequences in gravitational physics. This theory is studied to
comprehend its theoretical implications and importance in
astrophysical and cosmological evolution. This theory proposes that
the incorporation of non-metricity and the existence of matter
sources provide a more comprehensive representation of gravitational
connections.

This theory incorporates non-metricity in the gravitational action,
introducing additional degrees of freedom that result in novel
gravitational dynamics and cosmological solutions. Additionally, the
dependence on the matter-Lagrangian allows this theory to
effectively characterize the impact of matter composition on the
gravitational field. The objective of this modified theory is to
provide a comprehensive framework for gravitational physics and
cosmology, which can provide information about basic investigations
into the nature of gravity and the cosmos. Harko et al \cite{33}
investigated diverse cosmological applications by deriving evolution
equations and using particular functional forms for $f(\textsl{Q})$
gravity. Mandal and Sahoo \cite{34} analyzed the equation of state
($\textsl{EoS}$) parameter in the framework of non-minimally coupled
$f(\textsl{Q})$ gravity. Myrzakulov et al \cite{35} further
investigated the impact of bulk viscosity on late-time cosmic
acceleration in $f(\textsl{Q},\textsl{L}_m)$ gravity framework.

The accelerated expansion of the universe can be examined using
various \textsl{DE} models. Turner and White \cite{24} found that
the inflationary phase was not aligned with the present matter
density and they tackled this problem by introducing a
parameterization technique. Sahni et al \cite{25} introduced a set
of dimensionless diagnostic parameters to evaluate the
characteristics of \textsl{DE}. Chirde and Shekh \cite{27} used
\textsl{DE} model and  \textsl{EoS} parameter to examine the cosmic
acceleration in $f(\textsl{R},\textsl{T})$ gravity ($\textsl{T}$ is
the energy-momentum tensor). Arora et al \cite{29} explored
late-time cosmology involving dust matter in
$f(\textsl{Q},\textsl{T})$ framework. Solanki et al \cite{30}
examined the characteristics of various cosmic parameters in
$f(\textsl{Q})$ gravity to examine the dark universe. Mussatayeva et
al \cite{31} described various late-time cosmological phenomena in
the $f(\textsl{Q})$ gravity.

Reconstruction method in extended gravitational theories provide an
effective reflection to explain the current accelerated expansion.
Ebrahimi and Sheykhi \cite{38} examined the $\textsl{GDE}$ model in
the context of Brans-Dicke cosmology using the non-interaction
scenario. Saaidi et al \cite{39} used correspondence approaches to
reconstruct $f(\textsl{R})$ gravity models. Jawad \cite{41} studied
the evolutionary paths of few cosmic factors using the concept of
pilgrim $\textsl{DE}$ in $f(\mathbb{T},\mathbb{T}_\textsl{G})$
theory. Fayaz et al \cite{42} studied the $f(\textsl{R},\textsl{T})$
model in the context of the $\textsl{GDE}$ model and examined the
evolutionary picture of the universe. Sharif and Nawazish \cite{43}
modified the $f(\textsl{R})$ gravity model using the generalized
ghost pilgrim $\textsl{DE}$ model with Friedmann-Robertson-Walker
(\textsl{FRW}) universe. They discovered that the non-interacting
condition exhibits cosmic acceleration, which is characterized by
the presence of a positive curvature parameter. The progression of
the universe across matter and $\textsl{DE}$ eras has been studied
in \cite{52}. Zadeh et al \cite{53} found that the particle
perspective accurately characterizes the cosmic acceleration during
late time. Ghaffari et al \cite{54} used Tsallis generalized entropy
to establish a holographic $\textsl{DE}$ model in Brans-Dicke
Theory. Huang et al \cite{55} analyzed cosmic evolution through
Tsallis holographic $\textsl{DE}$ model. Odintsov et al \cite{37}
investigated distinct $f(\textsl{R},\textsl{G})$ models to elucidate
the successful manifestation of the $\textsl{DE}$ model. Myrzakulov
et al \cite{45} used the pilgrim $\textsl{DE}$ and $\textsl{GDE}$
models to rebuild the $f(\textsl{Q})$ model. They discovered that
the outcomes achieved closely reflect the observational data. The
construction of different models and cosmological analysis
corresponding to $\textsl{DE}$ models has been established in
\cite{49}-\cite{50}.

To examine the dynamics of the universe, we reconstruct the
$f(\textsl{Q}, \textsl{L}_{m})$ functional form using the
\textsl{GGDE} model to analyze the dynamics of the cosmos. This
article follows the given format. Section \textbf{2} presents the
basis of $f(\textsl{Q}, \textsl{L}_{m})$ gravity. Then, we assess
the influence of the connection among $\textsl{DE}$ and cold
$\textsl{DM}$ using analysis of the redshift parameter. We also
reconstruct $f(\textsl{Q}, \textsl{L}_{m})$ functional form by
considering $\textsl{GGDE}$ model. Section \textbf{3} focuses on
examining the evolution of this notion using cosmographic analysis.
We summarize our main findings in section \textbf{4}.

\section{$f(\textsl{Q},\textsl{L}_{m})$ Theory: Field Equations}

We derive the equations of motion of $f(\textsl{Q},\textsl{L}_{m})$
theory by variational principle in this section. A generalized form
of Riemannian geometry was presented by Weyl \cite{34} as a
mathematical foundation for the description of gravitation in
\textsl{GR}. In Riemannian geometry, parallel transport along a
closed path preserves the vector length and direction. Weyl
introduced a modification in which a vector undergoes changes in
both direction and length while being transported in parallel along
a closed path. The proposed modification introduces a novel vector
field $(\textsl{A}^{\xi})$ that defines the geometric features of
Weyl geometry. The primary fields in Weyl space consist of the newly
defined vector field and the metric tensor. The metric tensor
establishes the spatial arrangement of spacetime by specifying
distances and angles, while the vector field is introduced to
accurately represent the variation in length during parallel
transport. In a Weyl theory, the length changes as $\delta\ell=
\ell\textsl{A}_{\xi} \delta x^ {\xi}$ when a vector size $\ell$ is
transported with an infinitesimal path $\delta x^{\xi}$ \cite{35}.
This suggests that the change in the size of the vector is directly
related to the coefficient of connection, original length and the
displacement along the trajectory.

The variation in the vector length is expressed as $\delta\ell=
\ell\Upsilon_{\xi\eta}\delta h^{\xi\eta}$, where $\delta
h^{\xi\eta}$ is the area element and
\begin{equation}\label{1}
\Upsilon_{\xi\eta}=\nabla_{\eta}\textsl{A}_{\xi}-\nabla_{\xi}\textsl{A}_{\eta}.
\end{equation}
A spatial scaling length $\hat{\ell}=\upsilon(x)\ell$ transforms the
field equation $\hat{\textsl{A}}_{\xi}$ to
$\hat{\textsl{A}}_{\xi}=\textsl{A}_{\xi}+ (\ln\upsilon),_{\xi}$,
whereas the conformal transformations modify the elements of metric
tensor as $\hat{g}_{\xi\eta}=\upsilon^{2}g_{\xi\eta}$ and
$\hat{g}^{\xi\eta}= \upsilon^{-2}g^{\xi\eta}$, respectively. Among
the fundamental characteristics of Weyl geometry, a semi-metric
connection is defined as
\begin{equation}\label{2}
{\hat{\Gamma}}^{\vartheta}_{\xi\eta}=\Gamma^{\vartheta}_{\xi\eta}
+g_{\xi\eta}\textsl{A}^{\vartheta}-\delta^{\vartheta}_{\xi}\textsl{A}_{\eta}-
\delta^{\vartheta}_{\eta}\textsl{A}_{\xi},
\end{equation}
where Christoffel symbol is represented by
$\Gamma^{\vartheta}_{\xi\eta}$. The construction of a gauge
covariant derivative is possible by assuming that $
\hat{\Gamma}^{\vartheta}_{\xi\eta}$ is symmetric. Using the
covariant derivative, the expression for the Weyl curvature tensor
is
\begin{equation}\label{3}
\hat{\mathcal{S}}_{\xi\eta\vartheta\gamma}=\hat{\mathcal{S}}
_{(\xi\eta)\vartheta\gamma}+\hat{\mathcal{S}}_{[\xi\eta]\vartheta\gamma},
\end{equation}
where
\begin{equation}\nonumber
\hat{\mathcal{S}}_{[\xi\eta]\vartheta\gamma}=\mathcal{S}
_{\xi\eta\vartheta\gamma}+2\nabla_{\vartheta}\textsl{A}_{[\xi
g_{\eta}]\gamma}+2\nabla_{\gamma}\textsl{A}_{[\eta
g_{\xi}]\vartheta}+2\textsl{A}_{\vartheta}\textsl{A}_{[\xi
g_{\eta}]\vartheta}+2\textsl{A}_{\gamma}\textsl{A}_{[\eta
g_{\xi}]\vartheta}-2\textsl{A}^{2}g_{\vartheta[\xi g_{\eta}]\gamma}.
\end{equation}
After first contraction of the Weyl tensor, we have
\begin{eqnarray}\label{4}
\hat{\mathcal{S}}^{\xi}_{\eta}&=&\hat{\mathcal{S}}
^{\vartheta\xi}_{\vartheta\eta}=\mathcal{S}^{\xi}_{\eta}
+2\textsl{A}^{\xi}\textsl{A}_{\eta}+3\nabla_{\eta}\textsl{A}^{\xi}-\nabla_{\xi}\textsl{A}^{\eta}
+g^{\xi}_{\eta}(\nabla_{\vartheta}\textsl{A}^{\vartheta}-2\textsl{A}_{\vartheta}\textsl{A}^{\vartheta}).
\end{eqnarray}
Finally, the Weyl scalar is given by
\begin{equation}\label{5}
\hat{\mathcal{S}}=\bar{\mathcal{S}}^{\vartheta}_{\vartheta}=
\mathcal{S}+6(\nabla_{\xi}\textsl{A}^{\xi}-\textsl{A}_{\xi}\textsl{A}^{\xi}).
\end{equation}
Weyl-Cartan (\textsl{WC}) spaces, which include torsion offer an
expanded framework that goes beyond Riemannian and Weyl geometries.

The \textsl{WC} spacetimes are characterized by a symmetric metric
tensor defining the length of a vector and an asymmetric connection
determines the law of parallel transport as
$d\varsigma^{\xi}=-\varsigma^{\vartheta}\hat{\Gamma}^{\xi}_{\vartheta\eta}dx^{\eta}$
\cite{37}. The connection in this framework is given by
\begin{equation}\label{6}
\tilde{\Gamma}^{\vartheta}_{\xi\eta}={\Gamma}^{\vartheta}_{\xi\eta}
+\Psi^{\vartheta}_{\xi\eta}+\Omega^{\vartheta}_{\xi\eta}.
\end{equation}
Here, disformation tensor $(\Omega^{\vartheta}_{\xi\eta})$ and
contortion tensor $(\Psi^{\vartheta}_{\xi\eta})$ are expressed as
\begin{eqnarray}\label{7}
\Omega^{\vartheta}_{\xi\eta}&=&\frac{1}{2}g^{\vartheta\gamma}
(\textsl{Q}_{\xi\eta\gamma}
+\textsl{Q}_{\xi\eta\gamma}-\textsl{Q}_{\gamma\xi\eta}),
\\\label{8}
\Psi^{\vartheta}_{\xi\eta}&=&\tilde{\Gamma}^{\vartheta}_{[\xi\eta]}
+g^{\vartheta\gamma}g_{\xi\varepsilon}
\tilde{\Gamma}^{\varepsilon}_{[\eta\gamma]}+g^{\vartheta\gamma}
g_{\eta\varepsilon}\tilde{\Gamma}^{\varepsilon}_{[\xi\gamma]},
\end{eqnarray}
where
\begin{equation}\label{9}
\textsl{Q}_{\gamma\xi\eta}=\nabla_{\gamma}g_{\xi\eta}
=-g_{\xi\eta,\gamma}+g_{\eta\varepsilon}\hat{\Gamma}^{\varepsilon}_{\xi\gamma}
+g_{\varepsilon\xi}\hat{\Gamma}^{\varepsilon}_{\eta\gamma}.
\end{equation}
Here, \textsl{WC} connection is denoted by
$\tilde{\Gamma}^{\vartheta}_{\xi\eta}$. It is noted that the
\textsl{WC} geometry represents a specific case of Weyl geometry
when torsion is absent as verified from Eqs.(\ref{2}) and (\ref{6}),
where $\textsl{Q}_{\vartheta\xi\eta} =
-2g_{\xi\eta}\Psi_{\vartheta}$. Therefore, Eq.(\ref{6}) becomes
\begin{equation}\label{10}
\tilde{\Gamma}^{\vartheta}_{\xi\eta}={\Gamma}^{\vartheta}_{\xi\eta}
+g_{\xi\eta}\Psi^{\vartheta}
-\delta^{\vartheta}_{\xi}\Psi_{\eta}-\delta^{\vartheta}_{\eta}\Psi_{\xi}
+\Psi^{\vartheta}_{\xi\eta},
\end{equation}
with
\begin{equation}\label{11}
\Psi^{\vartheta}_{\xi\eta}=T^{\vartheta}_{\xi\eta}-g^{\vartheta\gamma}
g_{\varepsilon\xi}T^{\varepsilon}_{\gamma\eta}-g^{\vartheta\gamma}
g_{\varepsilon\eta}T^{\varepsilon}_{\gamma\xi}.
\end{equation}
The \textsl{WC} torsion is given by
\begin{equation}\label{12}
T^{\vartheta}_{\xi\eta}=\frac{1}{2}(\tilde{\Gamma}^{\vartheta}
_{\xi\eta}-\tilde{\Gamma}^{\vartheta}_{\eta\xi}).
\end{equation}
A connection-based definition of the \textsl{WC} curvature tensor is
\begin{equation}\label{13}
\tilde{\mathcal{S}}^{\vartheta}_{\xi\eta\gamma}=\tilde{\Gamma}^{\vartheta}
_{\xi\gamma,\eta}-\tilde{\Gamma}^{\vartheta}_{\xi\eta,\gamma}+\tilde{\Gamma}
^{\varepsilon}_{\xi\gamma}
\tilde{\Gamma}^{\vartheta}_{\varepsilon\eta}-\tilde{\Gamma}
^{\varepsilon}_{\xi\eta}
\tilde{\Gamma}^{\vartheta}_{\varepsilon\gamma}.
\end{equation}
Contraction of the curvature tensor yields the \textsl{WC} scalar as
\begin{eqnarray}\nonumber
\tilde{\mathcal{S}}&=&\tilde{\mathcal{S}}^{\xi\eta}_{\xi\eta}
=\mathcal{S}+6\nabla_{\eta}\Psi^{\eta}-4\nabla_{\eta}
T^{\eta}-6\Psi_{\eta}\Psi^{\eta} +8\textsl{A}
_{\eta}T^{\eta}+T^{\xi\vartheta\eta}T_{\xi\vartheta\eta}
\\\label{14}
&+&2T^{\xi\vartheta\eta}T_{\eta\vartheta\xi}-4T^{\eta}T_{\eta},
\end{eqnarray}
where $T_{\eta}=T^{\xi}_{\xi\eta}$.

By neglecting the boundary terms in the Ricci scalar, one can
reformulate the gravitational action as \cite{38}
\begin{equation}\label{15}
\textsl{S}=\frac{1}{2\kappa} \int
g^{\xi\eta}(\Gamma^{\vartheta}_{\gamma\xi}\Gamma^{\gamma}_{\vartheta\eta}
-\Gamma^{\vartheta}_{\gamma\vartheta}\Gamma^{\gamma}_{\xi\eta})\sqrt{-g}
d^{4}x.
\end{equation}
The assumption of symmetric connection yields
\begin{equation}\label{16}
\Gamma^{\vartheta}_{\xi\eta}=-\textsl{L}^{\vartheta}_{\xi\eta}.
\end{equation}
Thus, Eq.\eqref{15} turns out to be
\begin{equation}\label{17}
\textsl{S}=-\frac{1}{2\kappa} \int
g^{\xi\eta}(\textsl{L}^{\vartheta}_{\gamma\xi}\textsl{L}^{\gamma}_{\vartheta\eta}
-\textsl{L}^{\vartheta}_{\gamma\vartheta}\textsl{L}^{\gamma}_{\xi\eta})\sqrt{-g}
d^{4}x,
\end{equation}
where
\begin{equation}\label{18}
\textsl{Q}\equiv-g^{\xi\eta}(\textsl{L}^{\vartheta}_{\gamma\xi}\textsl{L}^{\gamma}_{\vartheta\eta}
-\textsl{L}^{\vartheta}_{\eta\vartheta}\textsl{L}^{\eta}_{\xi\eta}),
\end{equation}
and
\begin{equation}\label{19}
\textsl{L}^{\vartheta}_{\gamma\xi}\equiv-\frac{1}{2}g^{\vartheta\varepsilon}
(\nabla_{\xi}g_{\gamma\varepsilon}+\nabla_{\gamma}g_{\varepsilon\xi}
-\nabla_{\varepsilon}g_{\gamma\xi}).
\end{equation}
By substituting the non-metricity scalar with a general function in
Eq.(\ref{17}), the gravitational action of symmetric teleparallel
theory can be derived as
\begin{equation}\label{20}
\textsl{S}=\frac{1}{2\kappa}\int f(\textsl{Q})\sqrt{-g}d^{4}x.
\end{equation}

Now, we couple this action with the matter-Lagrangian density, the
action of $f(\textsl{Q},\textsl{L}_{m})$ theory is obtained as
\cite{6,Q}
\begin{equation}\label{21}
\textsl{S}=\frac{1}{2\kappa}\int f(\textsl{Q},\textsl{L}_{m})
\sqrt{-g}d^{4}x.
\end{equation}
The superpotential is given by
\begin{equation}\label{22}
\mathcal{P}^{\vartheta}_{\xi\eta}=-\frac{1}{2}\textsl{L}^{\vartheta}_{\xi\eta}
+\frac{1}{4}(\textsl{Q}^{\vartheta}
-\tilde{\textsl{Q}}^{\vartheta})g_{\xi\eta}- \frac{1}{4} \delta
^{\vartheta} _{[\xi \textsl{Q}_{\eta}]}.
\end{equation}
The relation for non-metricity (given in Appendix
\textbf{$\mathcal{X}$}) is
\begin{equation}\label{23}
\textsl{Q}=-\textsl{Q}_{\vartheta\xi\eta}\mathcal{P}
^{\vartheta\xi\eta}=-\frac{1}{4} (-\textsl{Q}^{\vartheta\xi\eta}
\textsl{Q}_{\vartheta\xi\eta}+2\textsl{Q}^{\vartheta\xi\eta}
\textsl{Q}_{\eta\vartheta\xi}
-2\textsl{Q}^{\vartheta}\tilde{\textsl{Q}}_{\vartheta}+\textsl{Q}
^{\vartheta}\textsl{Q}_{\vartheta}).
\end{equation}
The variation of Eq.(\ref{21}) gives
\begin{eqnarray}\nonumber
\delta\textsl{S}&=&\frac{1}{2}\int\delta
[f(\textsl{Q},\textsl{L}_{m}) \sqrt{-g}]d^{4}x,
\\\label{24}
&=&\frac{1}{2}\int(f\delta\sqrt{-g} +(f_{\textsl{Q}}\delta
\textsl{Q}+f_{\textsl{L}_{m}}\delta \textsl{L}_{m})\sqrt{-g}d^{4}x.
\end{eqnarray}
Moreover, we define
\begin{eqnarray}\label{25}
\textsl{T}_{\xi\eta}=-\frac{2}{\sqrt{-g}}\frac{\delta(\sqrt{-g}\textsl{L}_{m})}{\delta
g^{\xi\eta}}=g_{\xi\eta}\textsl{L}_{m}-2\frac{\partial
\textsl{L}_{m}}{\partial g^{\xi\eta}}.
\end{eqnarray}
The variation of $\textsl{Q}$ is discussed in Appendix
\textbf{$\mathcal{Y}$} and determinant of the metric tensor is given
by
\begin{equation}\label{26}
\delta\sqrt{-g}=-\frac{1}{2}\sqrt{-g}g_{\xi\eta}\delta g^{\xi\eta}.
\end{equation}
Using Eqs.\eqref{25}, \eqref{26} and variation of $\delta
\textsl{Q}$ in \eqref{24}, we have
\begin{eqnarray}\nonumber
\delta\textsl{S}&=&\frac{-1}{2}\int fg_{\xi\eta}\sqrt{-g}\delta
g^{\xi\eta}
\\\nonumber
&-&f_{\textsl{Q}}\sqrt{-g}(\mathcal{P}_{\xi\vartheta\gamma}
\textsl{Q}_{\eta}^{\vartheta\gamma}-2\textsl{Q}^{\vartheta\gamma}
_{\xi} \mathcal{P}_{\vartheta\gamma\eta}) \delta
g^{\xi\eta}+2f_{\textsl{Q}}\sqrt{-g}
\mathcal{P}_{\vartheta\xi\eta}\nabla^{\vartheta} \delta g^{\xi\eta}
\\\label{27}
&+&\frac{1}{2}f_{\textsl{L}_{m}}(\textsl{L}_{m}-\mathcal{T}_{\xi\eta})\sqrt{-g}
\delta g^{\xi\eta}d^{4}x.
\end{eqnarray}
Integrate while considering the boundary conditions, the term $2
f_{\textsl{Q}}\sqrt{-g}\mathcal{P}_{\vartheta\xi\eta}
\nabla^{\vartheta}\delta g^{\xi\eta}$ takes the form as $-2
\nabla^{\vartheta} (f_{\textsl{Q}} \sqrt{-g}
\mathcal{P}_{\vartheta\xi\eta})\delta g^{\xi\eta}$. The resulting
field equations of $f(\textsl{Q},\textsl{L}_{m})$ gravity are
\begin{eqnarray}\nonumber
\frac{1}{2}f_{\textsl{L}_{m}}(\textsl{L}_{m}-\textsl{T}_{\xi\eta})&=&
\frac{2}{\sqrt{-g}} \nabla_{\vartheta} (f_{\textsl{Q}}\sqrt{-g}
\mathcal{P}^{\vartheta}_{\xi\eta})+ \frac{1}{2}fg_{\xi\eta}
\\\label{28}
&+&f_{\textsl{Q}} (\mathcal{P}_{\xi\vartheta\gamma}
\textsl{Q}_{\eta}^{\vartheta\gamma}
-2\textsl{Q}^{\vartheta\gamma}_{\xi}
\mathcal{P}_{\vartheta\gamma\eta}),
\end{eqnarray}
where $f_{\textsl{L}_{m}}=\frac{\partial f}{\partial
\textsl{L}_{m}}$ and $f_{\textsl{Q}}=\frac{\partial f}{\partial
\textsl{Q}}$.

\subsection{Formulation of \textsl{GGDE} $f(\textsl{Q},\textsl{L}_{m})$ Model}

To explore the mysteries of the cosmos, we consider a homogeneous
and isotropic spacetime characterized by the scale factor
$\mathrm{a}(t)$ as
\begin{equation}\label{29}
ds^{2}=-dt^{2}+\mathrm{a}^{2}(t)(dx^{2}+dy^{2}+dz^{2}).
\end{equation}
We assume that the universe is filled with ideal fluid, whose
stress-energy tensor is expressed as
\begin{equation}\label{30}
\textsl{T}_{\xi\eta}=(\mu+\textsl{P})\mathcal{V}_{\xi}\mathcal{V}_{\eta}+\textsl{P}g_{\xi\eta},
\end{equation}
where $\mu$, $\textsl{P}$ and $\mathcal{V}_{\xi}$ represent the
energy density, pressure and four-velocity of the fluid,
respectively. The non-zero components of the non-metricity and
deformation tensor are
\begin{eqnarray}\nonumber
\textsl{Q}_{011}&=&\textsl{Q}_{022}=\textsl{Q}_{033}=2\mathrm{a}\dot{\mathrm{a}},
\\\nonumber
\textsl{Q}_{0}~^{11}&=&\textsl{Q}_{0}~^{22}=\textsl{Q}_{0}~^{33}=\frac{2\dot{\mathrm{a}}}{\mathrm{a}^{3}},
\\\nonumber
\textsl{Q}^{01}~_{1}&=&\textsl{Q}^{02}~_{2}=\textsl{Q}^{03}~_{3}=-\frac{2\dot{\mathrm{a}}}{\mathrm{a}},
\\\nonumber
\textsl{L}^{0}~_{11}&=&\textsl{L}^{0}~_{22}=\textsl{L}^{0}~_{33}=-\mathrm{a}\dot{\mathrm{a}},
\\\nonumber
\textsl{L}^{1}~_{01}&=&L^{1}~_{10}=\textsl{L}^{2}~_{02}=
\textsl{L}^{2}~_{20}=\textsl{L}^{3}~_{03}=\textsl{L}^{3}~_{30}=-\frac{\dot{\mathrm{a}}}{\mathrm{a}}.
\end{eqnarray}
Using the values of these non-zero components with flat
$\textsl{FRW}$ spacetime in Eq.\eqref{23}, we have
\begin{eqnarray}\nonumber
\textsl{P}^{0}~_{11}&=&\textsl{P}^{0}~_{22}=\textsl{P}^{0}~_{33}=-\mathrm{a}\dot{\mathrm{a}},
\\\nonumber
\textsl{P}^{011}&=&\textsl{P}^{022}=\textsl{P}^{033}=-\frac{\dot{\mathrm{a}}}{\mathrm{a}^{3}},
\\\nonumber
\textsl{P}_{011}&=&\textsl{P}_{022}=\textsl{P}_{033}=\mathrm{a}\dot{\mathrm{a}},
\\\nonumber
\textsl{P}^{1}~_{01}&=&\textsl{P}^{0}~_{10}=\textsl{P}^{2}~_{02}=
\textsl{P}^{2}~_{20}=\textsl{P}^{3}~_{03}=\textsl{P}^{3}~_{30}=-\frac{\dot{\mathrm{a}}}{4\mathrm{a}},
\\\nonumber
\textsl{P}_{110}&=&\textsl{P}_{101}=\textsl{P}_{220}=\textsl{P}_{202}=
\textsl{P}_{330}=\textsl{P}_{303}=-\frac{\mathrm{a}\dot{\mathrm{a}}}{4},
\\\nonumber
\textsl{P}^{110}&=&\textsl{P}^{101}=\textsl{P}^{220}=\textsl{P}^{202}=
\textsl{P}^{330}=\textsl{P}^{303}=-\frac{\dot{\mathrm{a}}}{4\mathrm{a}^{3}}.
\end{eqnarray}

The non-metricity scalar is calculated by using Eq.\eqref{25} as
\begin{equation}\nonumber
\textsl{Q}=-(\textsl{Q}_{011}\textsl{P}^{011}+\textsl{Q}_{022}\textsl{P}^{022}
+\textsl{Q}_{033}\textsl{P}^{033}).
\end{equation}
After simplifying this equation, we obtain
$\textsl{Q}=6\textsl{H}^{2}$, where
$\textsl{H}=\frac{\dot{\mathrm{a}}}{\mathrm{a}}$. Evaluating
Eq.\eqref{29} for 0-0 component, we have
\begin{equation}\nonumber
\frac{2}{\mathrm{a}^{3}}\nabla_{\mu}(f_{\textsl{Q}}\sqrt{-\mathrm{g}}\textsl{P}^{\mu}_{00})+f_{\textsl{Q}}
(\textsl{P}_{0\mu\nu}\textsl{Q}_{0}~^{\mu\nu}-2\textsl{Q}^{\mu\nu}~_{0}
\textsl{P}_{\mu\nu0})+\frac{1}{2}f
\mathrm{g}_{00}=\frac{1}{2}f_{\textsl{L}_{m}}(\mathrm{g}_{00}\textsl{L}_{m}-\textsl{T}_{00}).
\end{equation}
This equation turns out to be
\begin{equation}\nonumber
f_{\textsl{Q}}(\textsl{P}_{011}\textsl{Q}_{0}~^{11}
+\textsl{P}_{022}\textsl{Q}_{0}~^{22}+\textsl{P}_{033}\textsl{Q}_{0}~^{33})
-\frac{1}{2}f=\frac{1}{2}f_{\textsl{L}_{m}}(\rho+\textsl{L}_{m}).
\end{equation}
After simplification, we have
\begin{equation}\label{31}
3\textsl{H}^{2}=\frac{1}{4f_{\textsl{Q}}}\big(f-f_{\textsl{L}_{m}}(\mu+\textsl{L}_{m})\big).
\end{equation}
Equation \eqref{29} for the 1-1 component becomes
\begin{equation}\nonumber
\frac{2}{\mathrm{a}^{3}}\nabla_{\mu}(f_{\textsl{Q}}\sqrt{-\mathrm{g}}\textsl{P}^{\mu}~_{11})
+f_{\textsl{Q}}(\textsl{P}_{1\mu\nu}
\textsl{Q}_{1}~^{\mu\nu}-2\textsl{Q}^{\mu\nu}~_{1}\textsl{P}_{\mu\nu1})
+\frac{1}{2}f
\mathrm{g}_{11}=\frac{1}{2}f_{\textsl{L}_{m}}(\mathrm{g}_{11}\textsl{L}_{m}-\textsl{T}_{11}).
\end{equation}
Manipulation of this equation yields
\begin{equation}\nonumber
\frac{2}{\mathrm{a}^{3}}\frac{\partial}{\partial
t}(f_{\textsl{Q}}\mathrm{a}^{3}(-\mathrm{a}\dot{\mathrm{a}}))-4\dot{\mathrm{a}}^{2}
f_{\textsl{Q}}
+\frac{\mathrm{a}^{2}}{2}f=\frac{\mathrm{a}^{2}}{2}f_{\textsl{L}_{m}}(\textsl{L}_{m}-p).
\end{equation}
Rearranging this equation, we have
\begin{equation}\label{32}
\dot{\textsl{H}}+3\textsl{H}^{2}+\frac{\dot{f}_{\textsl{Q}}}
{f_{\textsl{Q}}}\textsl{H}=\frac{1}{4f_{\textsl{Q}}}\big(f+f_{\textsl{L}_{m}}(\mu-\textsl{L}_{m})\big),
\end{equation}
where dot represents the temporal derivative, $\mu_{D}$ and
$\textsl{P}_{D}$ represent the energy density and pressure
corresponding to \textsl{DE} expressed as
\begin{eqnarray}\label{33}
\mu_{D}&=&\frac{-12 \textsl{H}^2 f_{\textsl{Q}}-\textsl{L}_m
f_{\textsl{L}_m}+f}{f_{\textsl{L}}},
\\\label{34}
\textsl{P}_{D}&=&\frac{4(\dot{\textsl{L}}_m\textsl{H}f_{\textsl{Q}\textsl{L}}+3\textsl{H}^2(4\dot{\textsl{H}}
f_{\textsl{Q}\textsl{Q}}+f_{\textsl{Q}})+\dot{\textsl{H}}
f_{\textsl{Q}})+\textsl{L}_mf_{\textsl{L}_m})-f}{f_{\textsl{L}_m}}.
\end{eqnarray}

The relations for the fractional energy densities are specified as
follows
\begin{equation}\label{35}
\Omega_{D} =\frac{\mu_{D}}{3\textsl{H}^2}, \quad
\Omega_{m}=\frac{\mu_{m}}{3\textsl{H}^2},
\end{equation}
This implies that $1=\Omega_{D} +\Omega_{m}$, indicating the
interplay between $\textsl{DE}$ and $\textsl{DM}$. Consequently, the
conservation of energy densities for two fluids can be determined as
follows when they interact
\begin{eqnarray}\label{36}
\dot{\mu_{m}}+3\textsl{H}(\mu_{m}+\textsl{P}_{m}) =\Gamma, \quad
\dot{\mu_{D}}+3\textsl{H}(\mu_{D}+\textsl{P}_{D})=-\Gamma.
\end{eqnarray}
Here, $\Gamma$ is the interaction term. The interaction term must be
positive for energy transfer from \textsl{DE} to \textsl{DM}. In
this specific framework, we examine the equation $\Gamma=3\eta
\textsl{H}(\mu_m+\mu_D)=3\textsl{H}\eta \mu_D(1 + \beta)$, where
$\eta$ defines the coupling constant and $\beta$ is given by
\begin{equation}\label{37}
\beta=\frac{\mu_{m}}{\mu_{D}}=\frac{\Omega_{m}}{\Omega_{D}}=\frac{1-\Omega_{D}}{\Omega_{D}}.
\end{equation}
We can express the $\omega_{D}$ as
\begin{equation}\label{38}
\omega_{D} =-\frac{\frac{ 2 \eta }{\Omega }+1}{2-\Omega }.
\end{equation}
The dynamic \textsl{DE} models that incorporate direct relation
between energy density and the Hubble parameter is crucial for
elucidating the phenomenon of the accelerated expansion of the
universe. In this context, the energy density of the
$\mathcal{\textsl{GGDE}}$ model is represented as
\begin{equation}\label{39}
\mu_{D}=\alpha \textsl{H}+\beta \textsl{H}^2 .
\end{equation}
Using Eqs.\eqref{33} and \eqref{39}, we have
\begin{equation}\label{40}
\frac{-12 \textsl{H}^2 f_{\textsl{Q}}-\textsl{L}_m
f_{\textsl{L}}+f}{f_{\textsl{L}}}=\alpha \textsl{H}+\beta
\textsl{H}^2.
\end{equation}
We consider $\textsl{L}_m=p$ and use Eq.\eqref{36}, we obtain the
reconstructed \textsl{GGDE} $f(\textsl{Q},\textsl{L}_{m})$ model as
\begin{equation}\label{41}
f(\textsl{Q},\textsl{L}_{m})=-\frac{\alpha c_1\sqrt{\textsl{Q}}(\ln
(\textsl{Q})+2)}{2\sqrt{6}}-\frac{1}{3}\beta c_1 \textsl{Q},
\end{equation}
where $c_{1}$ is the integration constant. Substituting this
reconstructed functional form in Eqs.\eqref{33} and \eqref{34}, we
have
\begin{eqnarray}\label{42}
\mu_{D}&=&\frac{\sqrt{6}\alpha(6\textsl{H}^2(\ln(\textsl{Q})+2)-\textsl{Q}\ln
(\textsl{Q}))-2\beta\sqrt{\textsl{Q}}(\textsl{Q}-12
\textsl{H}^2)}{12\sqrt{\textsl{Q}}},
\\\nonumber
\textsl{P}_{D}&=&\frac{1}{12\textsl{Q}^{3/2}}(2\beta
\textsl{Q}^{3/2}(-4\dot{\textsl{H}}-12
\textsl{H}^2+\textsl{Q})+\sqrt{6}\alpha(\textsl{Q}(\textsl{Q}\ln(\textsl{Q})\\\label{43}
&-&2(\ln(\textsl{Q})+2)\dot{\textsl{H}})-6\textsl{H}^2(\textsl{Q}(\ln(\textsl{Q})+2)-2\ln(\textsl{Q})\dot{\textsl{H}}))).
\end{eqnarray}

Now, we examine the cosmic parameters through the redshift function,
which is significant to comprehend the dynamics and the evolution of
the universe. The dynamics of redshift provides insights on cosmic
acceleration and the motion of cosmic objects with time. In this
perspective, we consider the scale factor as
\begin{equation}\label{44}
\mathrm{a}(t)=\mathrm{a}_{0}t^{k},
\end{equation}
where $\mathrm{a}_{0}$ and $k$ are arbitrary constants. The
deceleration parameter is represented as
\begin{equation}\label{45}
\textsl{q}=-\frac{\mathrm{a}\ddot{\mathrm{a}}}{\dot{\mathrm{a}}^{2}}=-1+\frac{1}{k}.
\end{equation}
This is a crucial cosmographic parameter that demonstrates the rate
of the expanding universe, because the cosmos undergoes decelerated
expansion for positive values of deceleration parameter, whereas its
negative values signify accelerated cosmic expansion. By using the
value of $k$, Eq.\eqref{44} becomes
\begin{equation}\label{46}
\mathrm{a}(t)=t^{\frac{1}{1+\textsl{q}}},
\end{equation}
where the present deceleration parameter value is $\textsl{q} =
-0.832_{-0.091}^{+0.091}$ \cite{63}. The relation for $\textsl{H}$
and $\textsl{H}_{0}$ is represented as
\begin{equation}\label{47}
\textsl{H}=\frac{\dot{\mathrm{a}}}{\mathrm{a}}=\bigg(\frac{1}{1+\textsl{q}}\bigg)(\frac{1}{t}),
~~\textsl{H}_{0}=\bigg(\frac{1}{1+\textsl{q}}\bigg)(\frac{1}{t_{0}}).
\end{equation}
This signifies that the universe expansion is impacted by the
deceleration parameter and $\textsl{H}_{0}$. By computing the
correlation between the redshift parameter and the scale factor, we
have
\begin{eqnarray}\label{48}
\textsl{H}=\textsl{H}_{0}\mathcal{U}^{1+\textsl{q}}, \quad
\dot{\textsl{H}}=-\textsl{H}_{0}\mathcal{U}^{2+2\textsl{q}},
\end{eqnarray}
where $\mathcal{U}=1+z$ \cite{66}. The non-metricity term is given
by
\begin{equation}\label{49}
\textsl{Q}=6\textsl{H}_{0}^{2}\mathcal{U}^{2+2\textsl{q}}.
\end{equation}
We consider Substituting these values in Eq.\eqref{41}, we get
\begin{eqnarray}\nonumber
f(\textsl{Q},\textsl{L}_{m})&=&-\frac{1}{2}\alpha c_1
\sqrt{H_0^2\mathcal{U}^{2q+2}}(\log(6H_0^2\mathcal{U}^{2
q+2})+2)\\\label{50} &-&2\beta c_1H_0^2\mathcal{U}^{2q+2}.
\end{eqnarray}
\begin{figure}
\epsfig{file=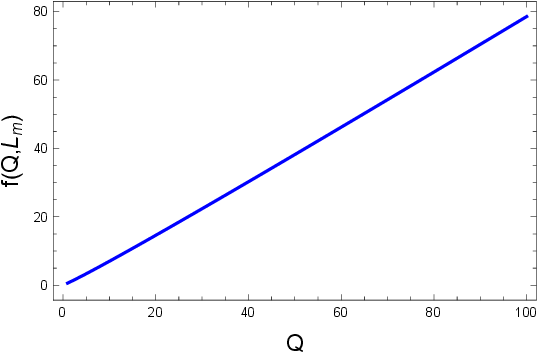,width=.5\linewidth}
\epsfig{file=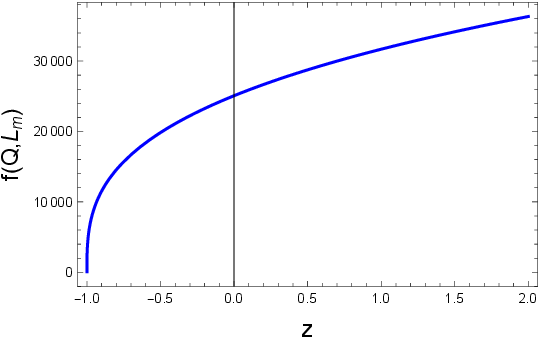,width=.5\linewidth}\caption{Graph of
$f(\textsl{Q},\textsl{L}_{m})$ versus non-metricity and redshift for
$\alpha=1.5$, $\beta=6.5$ and $c_{1}=0.4$.}
\end{figure}
Figure \textbf{1} demonstrates that the reconstructed
$f(\textsl{Q},\textsl{L}_{m})$ model retains a positive value and
exhibits an increasing trend corresponding to both redshift function
and non-metricity scalar. This graphical analysis indicates that the
reconstructed \textsl{GGDE} model supports the cosmic acceleration.
It is noteworthy that as $\textsl{Q}$ approaches to zero, our
rebuilt model converges to zero, signifying that the reconstructed
model exhibits realistic behavior. Substituting Eq.\eqref{49} in
\eqref{42} and \eqref{43}, we have
\begin{eqnarray}\label{51}
\mu_{D}&=&\alpha\sqrt{\textsl{H}_0^2\mathcal{U}^{2\textsl{q}+2}}+\beta
\textsl{H}_0^2\mathcal{U}^{2\textsl{q}+2},
\\\label{52}
\textsl{P}_{D}&=&-\alpha\sqrt{\textsl{H}_0^2\mathcal{U}^{2\textsl{q}+2}}-\beta
\textsl{H}_0^2\mathcal{U}^{2\textsl{q}+2}.
\end{eqnarray}
In cosmology, the composition of matter offers significant insights
into the cosmic evolution. Figure \textbf{2} determines the
graphical behavior of energy density and pressure corresponding to
reconstructed $\textsl{GGDE}$ $f(\textsl{Q},\textsl{L}_{m})$. As
observed from these plots, the pressure remains negative and the
energy density is positive. These graphical behaviors align with the
characteristics of $\textsl{DE}$, suggesting accelerated expansion.
The distinct behaviors of energy density and pressure help to
evaluate the viability of this bouncing model, underscoring its
relevance in advancing our understanding of cosmic evolution.
\begin{figure}
\epsfig{file=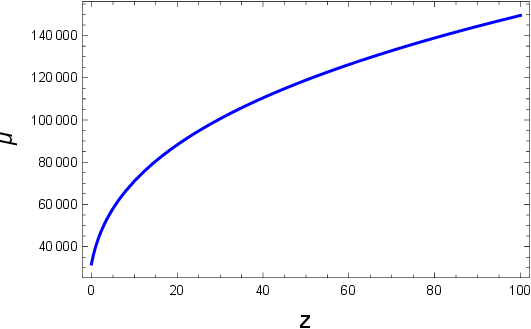,width=.5\linewidth}
\epsfig{file=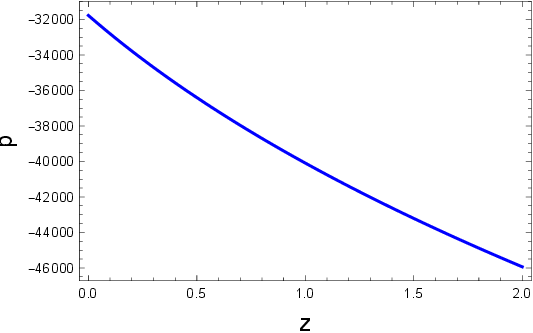,width=.5\linewidth}\caption{Graphical
representation of matter variables versus redshift for $\alpha=1.5$,
$\beta=6.5$ and $c_{1}=0.4$.}
\end{figure}

\section{Study of Cosmographic Parameters}

This section examines the dynamics of the cosmos through
cosmographic analysis of the key cosmic parameters for the
reconstructed $\textsl{GGDE}$ $f(\textsl{Q},\textsl{L}_{m})$
framework. Moreover, we examine consistency of this model using the
sound speed approach.

\subsection{Analysis of State Parameter}

The $\textsl{EoS}$ parameter $(\omega=\frac{\textsl{P}}{\mu})$
represents correlation between energy density and pressure. This
parameter helps to comprehend how these components impact cosmic
dynamics. This parameter sheds light on the forces driving cosmic
evolution and provides a more detailed understanding of the universe
progression. Different values of $\omega$ correspond to various
cosmic epochs, it distinguishes between different expansion phases,
where $\omega\in (-1,-\frac{1}{3})$ corresponds to quintessence era
and $\omega\in(-\infty,-1)$ describes the phantom phase. By applying
Eq.\eqref{38}, we get
\begin{equation}\label{53}
\omega_{D}=\frac{3(\sqrt{6}\alpha\sqrt{\textsl{Q}}+\textsl{Q}(\beta+6\eta
))}{6\alpha^2+2\sqrt{6}\alpha(\beta-3)\sqrt{\textsl{Q}}+(\beta
-6)\beta \textsl{Q}}.
\end{equation}
In terms of the redshift function, we have
\begin{equation}\label{54}
\omega_{D}=\frac{3\alpha\sqrt{\textsl{H}_0^2\mathcal{U}^{2\textsl{q}+2}}+3\textsl{H}_0^2(\beta
+6\eta)\mathcal{U}^{2\textsl{q}+2}}{(\alpha+(\beta-6)\sqrt{\textsl{H}_0^2
\mathcal{U}^{2\textsl{q}+2}})(\alpha+\beta\sqrt{\textsl{H}_0^2\mathcal{U}^{2
\textsl{q}+2}})}.
\end{equation}
Figure \textbf{3} shows the behavior of the \textsl{EoS} in the
\textsl{GGDE} $f(\textsl{Q},\textsl{L}_{m})$ theory for distinct
parametric values. This plot indicates that the \textsl{EoS}
parameter demonstrates a phantom regime, signifying that the cosmos
is undergoing accelerated cosmic expansion.
\begin{figure}\center
\epsfig{file=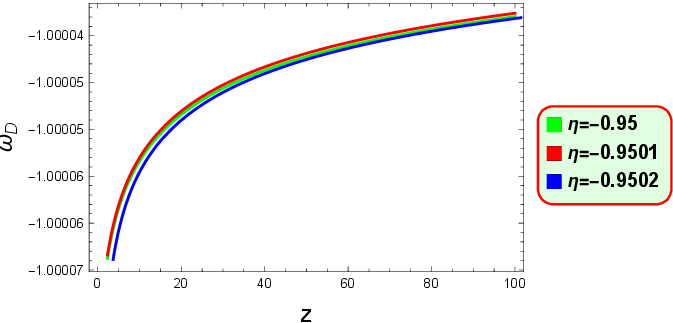,width=.5\linewidth}\caption{Graph of
$\omega_{D}$ versus redshift for $\alpha=1.5$, $\beta=6.5$ and
$c_{1}=0.4$.}
\end{figure}

\subsection{Examination of $(\omega_{D}-\omega'_{D})$-Plane}

Here, we employ ($\omega_{D}-\omega'_{D}$) analysis to examine the
dynamics of \textsl{DE}. This analysis elucidates the impact of the
modified terms on the deceleration parameter and the transition
among various cosmic phases. The dynamics of \textsl{DE} with the
scalar field has been studied in \cite{63}. Caldwell and Linder
\cite{66} classified \textsl{DE} frameworks into two different
categories, i.e., the thawing region and the freezing region. In the
thawing region, the cosmic acceleration happens over a short period
which is characterized by the positive value of $\omega'_{D}$ and
negative value of $\omega_D$. Whereas, the cosmic acceleration
happens over a long period in the frozen region, defined by negative
value of $\omega'_{D}$ and positive value of $\omega_D$. In the
($\omega_{D}-\omega'_{D}$) plane, the standard model is expressed by
the point $(-1,0)$. Using Eq.(\ref{53}), we have
\begin{equation}\label{55}
\omega'_{D}=\frac{3\alpha(6\sqrt{6}\alpha^2+12\alpha
\sqrt{\textsl{Q}}(\beta+6\eta)+\sqrt{6}\textsl{Q}(\beta^2+12(\beta-3)
\eta))}{2\sqrt{\textsl{Q}}(6\alpha^2+2\sqrt{6}\alpha
(\beta-3)\sqrt{\textsl{Q}}+(\beta-6)\beta \textsl{Q})^2}.
\end{equation}
This relation in terms of redshift function turns out to be
\begin{equation}\label{56}
\omega'_{D}=\frac{\alpha(\alpha(\alpha+2(\beta+6
\eta)\sqrt{\textsl{H}_0^2\mathcal{U}^{2\textsl{q}+2}})+\textsl{H}_0^2(\beta^2+12
(\beta-3)\eta)\mathcal{U}^{2\textsl{q}+2})}{4\sqrt{\textsl{H}_0^2\mathcal{U}^{2
\textsl{q}+2}}(\alpha(\alpha+2(\beta-3)\sqrt{\textsl{H}_0^2\mathcal{U}^{2
\textsl{q}+2}})+(\beta-6)\beta
\textsl{H}_0^2\mathcal{U}^{2\textsl{q}+2})^2}.
\end{equation}
Figure \textbf{4} exhibits that the values of $\omega_D$ and
$\omega'_{D}$ are in 0 and 1. This behavior corresponds to the
standard model, indicating that cosmic expansion is undergoing a
more accelerated rate in this framework.
\begin{figure}\center
\epsfig{file=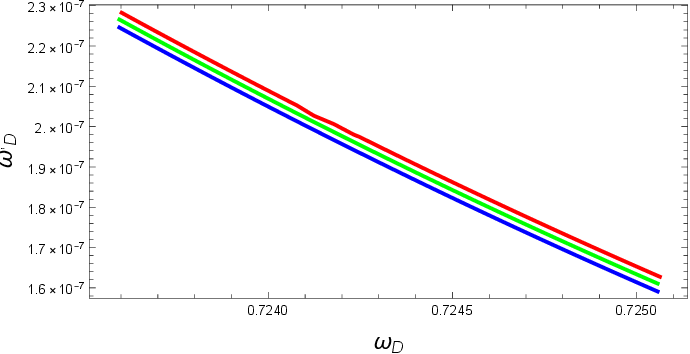,width=.5\linewidth}\caption{Plot of $\omega_{D}$
versus $\omega'_{D}$ for $\alpha=1.5$, $\beta=6.5$ and $c_{1}=0.4$.}
\end{figure}

\subsection{Investigation of $r-s$ Plane}

The $r-s$ parameters offer a more detailed comprehension of the
dynamical behavior and evolutionary phases of \textsl{DE} models
\cite{67}. These operators delineate the distinctions across diverse
cosmological models and also provide the distance of a specific
model through the lambda \textsl{CDM} limit. These two dimensionless
parameters characterize the standard model at $(r,s)=(1,1)$ and
\textsl{CDM} model at $(r,s)=(1,0)$. Furthermore, when the
trajectrories of the $(r-s)$plane fall in the interval of
$(r<1,s>0)$, one can experience the phantom and quintessence cosmic
eras. Conversely, the Chaplygin gas model is obtained for $r>1$ and
$s<0$. These are classified as \cite{68}
\begin{eqnarray}\label{57}
r&=&\frac{\dddot{\mathrm{a}}}{\mathrm{a}\textsl{H}^{3}}=
1+\frac{9\omega_{D}}{2}\Omega_{D}(1+\omega_{D})-\frac{3\omega'_{D}}{2\textsl{H}}\Omega_{D}
\\\nonumber
s&=&\frac{r-1}{3(\textsl{q}-\frac{1}{2})}=1+\omega_{D}-\frac{\omega'_{D}}{3\omega_{D}\textsl{H}}.
\end{eqnarray}
Substituting the values of $\omega_{D}$ and $\omega'_{D}$, we obtain
\begin{eqnarray}\nonumber
r&=&(-54\sqrt{6}\alpha^4+9\alpha
\textsl{Q}^{3/2}(44\alpha^3-\beta^3-12
(\beta-3)\beta\eta)+18\alpha^2\textsl{Q}^{5/2}(\beta(22\beta
\\\nonumber
&+&54\eta-51)+24)+\beta
\textsl{Q}^{7/2}(11\beta^3+\beta^2(54\eta-51)+72\beta+972\eta^2)\\\nonumber
&+&\sqrt{6}\alpha
\textsl{Q}^3(44\beta^3+9\beta^2(18\eta-17)+144\beta+972
\eta^2+6\sqrt{6}\alpha^3\textsl{Q}^2(44\beta\\\nonumber
&+&54\eta-51)-162
\alpha^3\sqrt{\textsl{Q}}(\beta+4\eta)-27\sqrt{6}\alpha^2\textsl{Q}(\beta^2+8\beta\eta-12\eta
)))\\\label{58}
&\times&(2\textsl{Q}^{3/2}(6\alpha^2+2\sqrt{6}\alpha(\beta-3)\sqrt{\textsl{Q}}+(\beta
-6) \beta \textsl{Q})^2)^{-1},\\\nonumber
s&=&(-6\alpha^3+6\sqrt{6}\alpha^3
\textsl{Q}^{3/2}+3\sqrt{6}\alpha\beta
\textsl{Q}^{5/2}(\beta+4\eta-2)+\textsl{Q}^3 (\beta +6
\eta)\\\nonumber
&\times&((\beta-3)\beta+18\eta)+18\alpha^2\textsl{Q}^2(\beta+2\eta-1)-2\sqrt{6}
\alpha^2\sqrt{\textsl{Q}}(\beta+6\eta)\\\nonumber&-&\alpha
\textsl{Q}(\beta^2+12(\beta-3)\eta
))(\textsl{Q}^{3/2}(6\alpha^2+2\sqrt{6}\alpha(\beta-3)\sqrt{\textsl{Q}}+(\beta-6)
\beta \textsl{Q})
\\\label{59} &\times&
(\sqrt{6}\alpha+\sqrt{\textsl{Q}}(\beta+6\eta)))^{-1}.
\end{eqnarray}
These parameters in terms of the redshift function are
\begin{eqnarray}\nonumber
r&=&(-\alpha^4+\alpha(44\alpha^3-\beta^3-12(\beta-3)\beta
\eta)(\textsl{H}_0^2\mathcal{U}^{2\textsl{q}+2})^{3/2}-3\alpha^3(\beta+4\eta)\\\nonumber
&\times&\sqrt{\textsl{H}_0^2\mathcal{U}^{2
\textsl{q}+2}}+4\alpha^3\textsl{H}_0^4(44\beta+54\eta-51)\mathcal{U}^{4
\textsl{q}+4}-3 \alpha^2\textsl{H}_0^2(\beta^2+8\beta\eta\\\nonumber
&-&12\eta)\mathcal{U}^{2\textsl{q}+2}+12\alpha^2
(\beta(22\beta+54\eta-51)+24)(\textsl{H}_0^2\mathcal{U}^{2\textsl{q}+2})^{5/2}\\\nonumber
&+&4\alpha
\textsl{H}_0^6(44\beta^3+9\beta^2(18\eta-17)+144\beta+972\eta^2)
\mathcal{U}^{6 \textsl{q}+6}+4\beta(11\beta^3\\\nonumber
&+&\beta^2(54\eta-51)+72\beta+972\eta^2)(\textsl{H}_0^2
\mathcal{U}^{2 \textsl{q}+2})^{7/2})(8(\textsl{H}_0^2\mathcal{U}^{2
\textsl{q}+2})^{3/2}\\\label{60}
&\times&(\alpha(\alpha+2(\beta-3)\sqrt{\textsl{H}_0^2 \mathcal{U}^{2
\textsl{q}+2}})+(\beta-6)\beta
\textsl{H}_0^2\mathcal{U}^{2\textsl{q}+2})^2)^{-1}\\\nonumber
s&=&(\alpha \textsl{H}_0^2\mathcal{U}^{2
\textsl{q}+2}(36(\eta+\alpha^2\sqrt{\textsl{H}_0^2 \mathcal{U}^{2
\textsl{q}+2}})-\beta(\beta+12\eta))-\alpha^2(\alpha\\\nonumber
&+&2(\beta+6\eta)\sqrt{\textsl{H}_0^2\mathcal{U}^{2\textsl{q}+2}})+108\alpha
\textsl{H}_0^4\mathcal{U}^{4\textsl{q}+4}(\alpha(\beta
+2\eta-1)\\\nonumber
&+&\beta(\beta+4\eta-2)\sqrt{\textsl{H}_0^2\mathcal{U}^{2
\textsl{q}+2}})+36\textsl{H}_0^6
(\beta+6\eta)((\beta-3)\beta+18\eta)\\\nonumber
&\times&\mathcal{U}^{6\textsl{q}+6})(36(\textsl{H}_0^2\mathcal{U}^{2
\textsl{q}+2})^{3/2}(\alpha(\alpha+2(\beta-3)\sqrt{\textsl{H}_0^2
\mathcal{U}^{2 \textsl{q}+2}})\\\label{61} &+&(\beta-6)\beta
\textsl{H}_0^2\mathcal{U}^{2 \textsl{q}+2})(\alpha+(\beta+6\eta)
\sqrt{\textsl{H}_0^2\mathcal{U}^{2 \textsl{q}+2}}))^{-1}.
\end{eqnarray}
Figure \textbf{5} demonstrates that $r>1$ and $s<0$, indicating the
Chaplygin gas model.
\begin{figure}\center
\epsfig{file=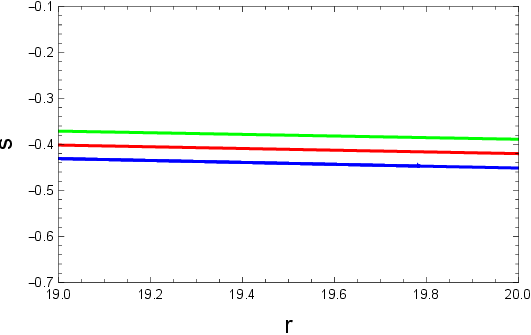,width=.5\linewidth}\caption{Plot of $r$ against
$s$ corresponding to $\alpha=1.5$, $\beta=6.5$ and $c_{1}=0.4$.}
\end{figure}

\subsection{Stability Analysis}

In the realm of modified gravity theories, stability analysis plays
a crucial role in determining the physical viability of any
cosmological model. A critical aspect of this investigation is to
ensure that the model remains stable under small perturbations, thus
avoiding instabilities. One key parameter in this analysis is the
squared sound speed $(v_{s}^{2})$, which characterizes the
propagation of perturbations in the cosmic fluid. A physically
viable model must maintain $v_{s}^{2}\geq 0$ to avoid the
irregularities, which lead to instabilities such as growing
perturbations or shock waves. In the background of $f(\textsl{Q},
\textsl{L}_{m})$ combined with $\textsl{GGDE}$ model, the
interaction between matter, non-metricity and $\textsl{GGDE}$
introduce new terms affecting $(v_{s}^{2})$, making it crucial to
ensure the model remains stable against perturbations. Stability
analysis, including the behavior of the $(v_{s}^{2})$, allows us to
explore the ranges of model parameters where perturbations decay and
the evolution remains physically consistent with cosmic
observations. This analysis reveal whether the $f(\textsl{Q},
\textsl{L}_{m})$ along with $\textsl{GGDE}$ model offers solutions
that remain stable throughout the cosmic evolution, particularly
during the transition from early cosmic inflation to the current
accelerated expansion.

By ensuring the stability of this extended gravity framework,
including the positivity of the $(v_{s}^{2})$, we apply the model to
address key cosmological questions, such as the nature of
$\textsl{GGDE}$. This analysis not only solidifies the theoretical
foundation of the $f(\textsl{Q}, \textsl{L}_{m})$ in the framework
of $\textsl{GGDE}$ but also positions it as a promising alternative
to standard cosmology. This parameter is given by
\begin{eqnarray}\nonumber
v_{s}^{2}&=&\frac{{\textsl{P}'}_{D}}{{\mu'}_{D}}
=\frac{\mu_{D}}{{\mu'}_{D}}\omega'_{D}+\omega_{D}\\\nonumber
&=&(6\sqrt{\textsl{Q}}(36\alpha^4+\sqrt{6}\alpha\beta
\textsl{Q}^{3/2}(\beta(4
\beta+21\eta-15)-72\eta)+(\beta-6)\beta^2\textsl{Q}^2(\beta\\\nonumber
&+&6\eta)+6\sqrt{6}\alpha^3\sqrt{\textsl{Q}}(4\beta+9\eta-3)+36\alpha^2\textsl{Q}(\beta(\beta+4\eta-2)-6
\eta)))\\\label{62}&\times&((\sqrt{6}\alpha
+2\beta\sqrt{\textsl{Q}})(6\alpha^2+2\sqrt{6}\alpha
(\beta-3)\sqrt{\textsl{Q}}+(\beta-6)\beta \textsl{Q})^2)^{-1}.
\end{eqnarray}
Substituting the value of non-metricity, we have
\begin{eqnarray}\nonumber
v_{s}^{2}&=&(6\sqrt{\textsl{H}_0^2\mathcal{U}^{2
\textsl{q}+2}}(\alpha^3(\alpha+(4
\beta+9\eta-3)\sqrt{\textsl{H}_0^2\mathcal{U}^{2\textsl{q}+2}})\\\nonumber
&+&\alpha \textsl{H}_0^2\mathcal{U}^{2\textsl{q}+2}(6
\alpha(\beta(\beta+4\eta-2)-6\eta)+\beta(\beta(4\beta+21\eta-15)\\\nonumber
&-&72
\eta)\sqrt{\textsl{H}_0^2\mathcal{U}^{2\textsl{q}+2}})+(\beta-6)\beta^2\textsl{H}_0^4
(\beta+6\eta)\mathcal{U}^{4\textsl{q}+4}))\\\nonumber
&\times&((\alpha+2\beta\sqrt{\textsl{H}_0^2\mathcal{U}^{2\textsl{q}+2}})
(\alpha(\alpha+2(\beta-3)\\\label{63}
&\times&\sqrt{\textsl{H}_0^2\mathcal{U}^{2\textsl{q}+2})+(\beta-6)\beta
\textsl{H}_0^2 \mathcal{U}^{2 \textsl{q}+2})^2})^{-1}.
\end{eqnarray}
Figure \textbf{6} demonstrates that the $(v_{s}^{2})>0$ in the
framework of $f(\textsl{Q}, \textsl{L}_{m})$ gravity for the
$\textsl{GGDE}$ model. This result is significant for the viability
and robustness of the model as it indicates that the model remains
well-behaved under perturbations and aligns with fundamental
stability criteria in cosmological evolution. In this context, the
$f(\textsl{Q}, \textsl{L}_{m})$ combined with $\textsl{GGDE}$ model
offers an enriched framework, potentially explaining the cosmic
accelerated expansion. The stability provided by $v_{s}^{2}>0$
reinforces the model capacity to address large-scale cosmological
observations while maintaining internal consistency. This analysis
not only affirms the stability of the cosmic evolution for the
$\textsl{GGDE}$ model but also supports the broader applicability of
the model in explaining key features of the universe expansion. The
stable propagation of perturbations indicates that the model can be
used to explore a wide range of cosmological phenomena from early
universe inflation to late-time cosmic acceleration, offering
valuable insights into the interaction between $\textsl{DE}$ and
modified gravity. Thus, the positivity of the squared sound speed
serves as a cornerstone for the reliability and physical
applicability.
\begin{figure}\center
\epsfig{file=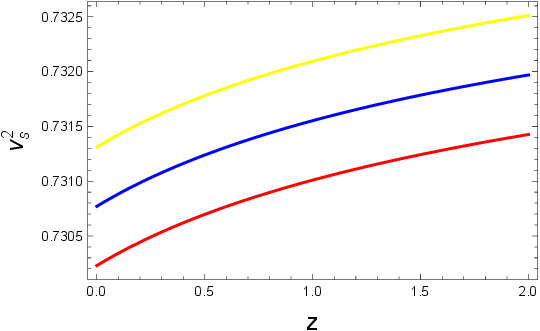,width=.5\linewidth}\caption{Graph of squared
sound speed versus redshift function for $\alpha=1.5$, $\beta=6.5$
and $c_{1}=0.4$.}
\end{figure}

\section{Summary and Discussion}

The reconstruction process in modified theory provides a useful
mechanism for developing a feasible \textsl{DE} model, capable of
precisely predicting the trajectory of cosmic evolution. The main
objective to explore a modified $f(\textsl{Q},\textsl{L}_{m})$
theory stems from numerous critical reasons pertaining to both
theoretical and observational frameworks in cosmology and
gravitational physics. This modified framework explains the cosmic
acceleration without the need for a cosmological constant. The
structure of the $f(\textsl{Q},\textsl{L}_{m})$ theory is inspired
by other established modified theories. The combination of
non-metricity and the matter-source term facilitates the development
of a geometrically coherent theory that preserves essential
phenomenological advantages, including the elucidation of late-time
acceleration and producing viable inflationary models. Thus, this
alternative theory aims to transcend the constraints of \textsl{GR}
to address cosmological singularities and investigate innovative
geometric structures via non-metricity, offering a cohesive
explanation for \textsl{DE}, \textsl{DM} and cosmic inflation. The
inclusion of the matter-source term enriches the theoretical
framework, offering new avenues for cosmological model building and
consistency with current observations.

In this manuscript, we have explored the novel cosmological insights
of $f(\textsl{Q},\textsl{L}_{m})$ theory by considering
\textsl{GGDE} model. The diagnostic tools and the statefinder pair
are used to analyze the various eras of the cosmos. Furthermore, the
model stability is evaluated using the squared sound speed method.
The significant results are detailed below.
\begin{itemize}
\item
The reconstructed \textsl{GGDE} $f(\textsl{Q},\textsl{L}_{m})$
gravity exhibits a rising trend corresponding to non-metricity and
redshift function, signifying the realistic behavior of the model
(Figure \textbf{1}).
\item
The increasing behavior of energy density and decreasing pattern of
pressure implies that the current cosmos is in a state of expansion
(Figure \textbf{2}).
\item
Our analysis also reveals that the \textsl{EoS} parameter describes
the standard cosmic model, aligning with the observed behavior of
the universe (Figure \textbf{3}).
\item
The increasing trend in the ($\omega_D-\omega'_D$) plane indicates
the freezing zone corresponding to different values of $\eta$
(Figure \textbf{4}). This suggests that \textsl{GGDE} theory results
in the rapid expansion of the cosmos.
\item
The $(r-s)$ plane demonstrates the Chaplygin gas model, indicating
that the reconstructed functional form support the cosmic
acceleration (Figure \textbf{5}).
\item
Our findings indicate that the reconstructed \textsl{GGDE}
$f(\textsl{Q},\textsl{L}_{m})$ gravity is stable as squared sound
speed is positive (Figure \textbf{6}).
\end{itemize}

The findings of our study significantly advance the understanding of
\textsl{DE} models by providing a comprehensive framework that links
\textsl{GGDE} with the extended $f(\textsl{Q},\textsl{L}_{m})$
gravitational theory. By reconstructing the
$f(\textsl{Q},\textsl{L}_{m})$ model in the context of
\textsl{GGDE}, our research reveals a nuanced interplay between
non-metricity, matter sources and cosmic evolution. This approach
offers a novel perspective on the theoretical underpinnings of dark
energy, particularly by addressing limitations in the $\Lambda$CDM
model. Unlike standard models, the reconstructed
$f(\textsl{Q},\textsl{L}_{m})$ framework incorporates both geometric
and matter-Lagrangian terms, providing a more holistic depiction of
the cosmos. Our findings demonstrate that the reconstructed
\textsl{GGDE} model aligns well with recent observational data,
reported by the Planck satellite and the Hubble parameter derived
from CMB studies. Notably, the diagnostic tools such as the
$(\omega_D, \omega'_D)$-plane and the $(r, s)$-plane, show that the
model captures the freezing regime of \textsl{DE}, indicating of an
accelerated expansion phase consistent with observational data.
Furthermore, the analysis of the energy density and pressure
evolution underlines a phantom-like behavior, signifying a more
rapid expansion of the cosmos and offering insights into possible
late-time cosmic scenarios. These results provide robust evidence
that the $f(\textsl{Q},\textsl{L}_{m})$ framework is a viable
alternative to \textsl{DE} models, offering compatibility with
observational constraints while introducing new pathways to explore
the dynamics of the dark universe. By bridging theoretical
constructs with empirical data, this study not only enriches the
field of modified gravity but also sets the stage for further
exploration of non-metricity-driven cosmic phenomena, making a
substantial contribution to understanding the nature of \textsl{DE}
and the accelerated expansion of the universe.

The \textsl{GGDE} $f(\textsl{Q},\textsl{L}_{m})$ model has stable
properties and consistently adheres to the current cosmic
accelerated expansion. The phantom nature of the cosmos is observed
to indicate a more rapid regime, potentially resulting in the
current cosmic acceleration. The results align with the existing
observational data given as \cite{69}
\begin{itemize}
\item
$\omega_{D}=-1.023_{-0.096}^{+0.091}$~~(Planck TT+LowP+ext),
\item
$\omega_{D}=-1.006_{-0.091}^{+0.085}$~~(Planck TT+LowP+lensing+ext),
\item
$\omega_{D}=-1.0019_{-0.080}^{+0.075}$~~(Planck TT, TE,
EE+LowP+ext).
\end{itemize}
The data has been obtained using various observational methods with
a confidence level of $95\%$. Our findings align with the
\textsl{DE} model in $f(\textsl{Q})$ \cite{70} and
$f(\textsl{Q},\textsl{T})$ gravity \cite{70a}. It is worthwhile to
highlight that our findings also align with the most recent data
from theoretical observations \cite{68}. Sharif and Zubair
\cite{70b} delved into the evolution of pilgrim \textsl{DE}
corresponding to event horizon, particle horizon and conformal age
of the universe in the framework of the \textsl{FRW} universe but
they did not check stability. We have also analyzed the stability of
the system. Our findings are found to be more concise with
observational data.

\vspace{0.25cm}

\section*{Appendix $\mathcal{X}$: Computation of \textsl{Q}}

\renewcommand{\theequation}{$\mathcal{X}$\arabic{equation}}
\setcounter{equation}{0}

Using Eqs.(\ref{19}) and (\ref{20}), we obtain
\begin{eqnarray}\label{A1}
\textsl{Q}&\equiv&-\mathrm{g}^{\xi\eta}\big(\textsl{L}^{\gamma}_{~\vartheta\xi}
\textsl{L}^{\vartheta}_{~\eta\gamma}-
\textsl{L}^{\gamma}_{~\vartheta\gamma}\textsl{L}^{\vartheta}_{~\xi\eta}\big),
\\\label{A2}
\textsl{L}^{\gamma}_{~\vartheta\xi}&=&-\frac{1}{2}\mathrm{g}^{\gamma\varepsilon}
\big(\textsl{Q}_{\xi\vartheta\varepsilon}+
\textsl{Q}_{\vartheta\varepsilon\xi}-\textsl{Q}_{\varepsilon\vartheta\xi}\big),
\\\label{A3}
\textsl{L}^{\vartheta}_{~\eta\gamma}&=&-\frac{1}{2}\mathrm{g}^{\vartheta\psi}
\big(\textsl{Q}_{\gamma\eta\psi}+
\textsl{Q}_{\eta\gamma\psi}-\textsl{Q}_{\psi\eta\gamma}\big),
\\\label{A4}
\textsl{L}^{\gamma}_{~\vartheta\gamma}&=&-\frac{1}{2}\mathrm{g}^{\gamma
\psi} \big(\textsl{Q}_{\gamma\vartheta\psi}+
\textsl{Q}_{\vartheta\psi\gamma
}-\textsl{Q}_{\psi\gamma\vartheta}\big),
\\\label{A5}
\textsl{L}^{\vartheta}_{~\xi\eta}&=&-\frac{1}{2}\mathrm{g}^{\vartheta\psi}
\big(\textsl{Q}_{\eta\xi\psi}+
\textsl{Q}_{\xi\psi\eta}-\textsl{Q}_{\psi\xi\eta} \big).
\end{eqnarray}
Therefore, we get
\begin{eqnarray}\label{A6}
-\mathrm{g}^{\xi\eta}\textsl{L}^{\gamma
}_{~\vartheta\xi}\textsl{L}^{\vartheta}_{~\eta\gamma}
&=&-\frac{1}{4}\big(2\textsl{Q}^{\gamma\eta\psi}
\textsl{Q}_{\psi\gamma\eta}-\textsl{Q}^{\gamma\eta\psi}
\textsl{Q}_{\gamma\eta\psi}\big),
\\\label{A7}
\mathrm{g}^{\xi\eta}\textsl{L}^{\gamma}_{~\vartheta\gamma}\textsl{L}^{\vartheta}_{~\xi\eta}
&=&\frac{1}{4}\mathrm{g}^{\xi\eta}\mathrm{g}^{\vartheta\varepsilon}\textsl{Q}_\vartheta
\big(\textsl{Q}_{\eta\xi\varepsilon}+\textsl{Q}_{\xi\varepsilon\eta}
-\textsl{Q}_{\varepsilon\eta\xi}\big),
\\\label{A8}
\textsl{Q}&=&-\frac{1}{4}\big(-\textsl{Q}^{\gamma\eta\xi}
\textsl{Q}_{\gamma\eta\xi}
+2\textsl{Q}^{\gamma\eta\xi}\textsl{Q}_{\xi\gamma\eta}
-2\textsl{Q}^{\gamma}\tilde{\textsl{Q}}_{\gamma}
+\textsl{Q}^{\gamma}\textsl{Q}_{\gamma}\big).
\end{eqnarray}
Using Eq.(\ref{23}), we have
\begin{eqnarray}\nonumber
\textsl{P}^{\gamma\xi\eta}&=&\frac{1}{4}\bigg[-\textsl{Q}^{\gamma\xi\eta}
+\textsl{Q}^{\xi\gamma\eta}+\textsl{Q}^{\eta\gamma\xi}
+\textsl{Q}^{\gamma}\mathrm{g}^{\xi\eta}-\tilde{\textsl{Q}}^{\gamma}
\mathrm{g}^{\xi\eta}-
\frac{1}{2}(\mathrm{g}^{\gamma\xi}\textsl{Q}^{\eta}
\\\label{A9}
&+&\mathrm{g}^{\gamma\eta}\textsl{Q}^{\xi})\bigg],
\\\label{A10}
-\textsl{Q}_{\gamma\xi\eta}\textsl{P}^{\gamma\xi\eta}
&=&-\frac{1}{4}\big(-\textsl{Q}^{\gamma\xi\eta}
\textsl{Q}_{\gamma\xi\eta}+2\textsl{Q}_{\gamma\xi\eta}
\textsl{Q}^{\xi\gamma\eta} +\textsl{Q}^{\gamma}\textsl{Q}_{\gamma}
-2\textsl{Q}_{\gamma}\tilde{\textsl{Q}}^{\gamma}\big) =\textsl{Q}.
\end{eqnarray}

\section*{Appendix $\mathcal{Y}$: \textbf{Variation of Q}}

\renewcommand{\theequation}{$\mathcal{Y}$\arabic{equation}}
\setcounter{equation}{0}

We consider Q as
\begin{eqnarray}\label{B1}
\textsl{Q}_{\gamma\xi\eta}&=&\nabla_{\gamma}\mathrm{g}_{\xi\eta},
\\\label{B2}
\textsl{Q}^{\gamma}_{~\xi\eta}&=&\mathrm{g}^{\gamma\vartheta}
\textsl{Q}_{\vartheta\xi\eta}=
\mathrm{g}^{\gamma\vartheta}\nabla_{\vartheta}\mathrm{g}_{\xi\eta}=\nabla^{\gamma}
\mathrm{g}_{\xi\eta},
\\\label{B3}
\textsl{Q}^{~~\xi}_{\gamma~~\eta}&=&\mathrm{g}^{\xi\vartheta}
\textsl{Q}_{\gamma\vartheta\eta}=\mathrm{g}^{\xi\vartheta}\nabla_{\gamma}
\mathrm{g}_{\vartheta\eta}=-\mathrm{g}_{\vartheta\eta}\nabla_{\gamma}\mathrm{g}^{\xi\vartheta},
\\\label{B4}
\textsl{Q}^{~~\eta}_{\gamma\xi}&=&\mathrm{g}^{\eta\vartheta}
\textsl{Q}_{\gamma\xi\vartheta}=\mathrm{g}^{\eta\vartheta}\nabla_{\gamma}
\mathrm{g}_{\xi\vartheta}=-\mathrm{g}_{\xi\vartheta}\nabla_{\gamma}\mathrm{g}^{\eta\vartheta},
\\\label{B5}
\textsl{Q}^{\gamma\xi}_{~~\eta}&=&\mathrm{g}^{\gamma\vartheta}
\mathrm{g}^{\xi\varepsilon}\nabla_{\vartheta}\mathrm{g}_{\varepsilon\eta}
=\mathrm{g}^{\xi\varepsilon}\nabla^{\gamma}\mathrm{g}_{\varepsilon\eta}=-\mathrm{g}_{\varepsilon\eta}
\nabla^{\gamma}\mathrm{g}^{\xi\varepsilon},
\\\label{B6}
\textsl{Q}^{\gamma~\eta}_{~\xi}&=&\mathrm{g}^{\gamma\vartheta}
\mathrm{g}^{\eta\varepsilon}\nabla_{\vartheta}\mathrm{g}_{\xi\varepsilon}
=\mathrm{g}^{\eta\varepsilon}\nabla^{\gamma}\mathrm{g}_{\xi\varepsilon}=-\mathrm{g}_{\xi\varepsilon}
\nabla^{\gamma}\mathrm{g}^{\eta\varepsilon},
\\\label{B7}
\textsl{Q}^{~~\xi\eta}_{\gamma}&=&\mathrm{g}^{\xi\varepsilon}\mathrm{g}^{\eta
\vartheta}\nabla_{\gamma}\mathrm{g}_{\varepsilon\vartheta}
=-\mathrm{g}^{\xi\varepsilon}\mathrm{g}_{\varepsilon\vartheta}\nabla_{\gamma}\mathrm{g}^{\eta\vartheta}
=-\nabla_{\gamma}\mathrm{g}^{\xi\eta},
\\\label{B8}
\textsl{Q}^{\gamma\xi\eta}&=&-\nabla^{\gamma}\mathrm{g}_{\xi\eta}.
\end{eqnarray}
Using Eqs.(\ref{B6}) and (\ref{B7}), we obtain
\begin{eqnarray}\nonumber
\delta \textsl{Q}&=&-\frac{1}{4}\delta\bigg(-\textsl{Q}
^{\gamma\eta\xi} \textsl{Q}_{\gamma\eta\xi}
+2\textsl{Q}^{\gamma\eta\xi}\textsl{Q}_{\xi\gamma\eta}
-2\textsl{Q}^{\gamma}\tilde{\textsl{Q}}_{\gamma}
+\textsl{Q}^{\gamma}\textsl{Q}_{\gamma}\bigg),
\\\nonumber
&=&-\frac{1}{4}\bigg(-\delta \textsl{Q}^{\gamma \eta\xi}
\textsl{Q}_{\gamma\eta\xi} -\textsl{Q}^{\gamma\eta\xi}\delta
\textsl{Q}_{\gamma\eta\xi} +2\delta \textsl{Q}^{\gamma
\eta\xi}\textsl{Q}_{\xi\gamma\eta}
\\\nonumber
&+&2\textsl{Q}^{\gamma \eta\xi}\delta \textsl{Q}_{\xi\gamma\eta}
-2\delta \textsl{Q}^{\gamma}\tilde{\textsl{Q}}_{\gamma}
-2\textsl{Q}^{\gamma}\delta\tilde{\textsl{Q}}_{\gamma} +\delta
\textsl{Q}^{\gamma}\textsl{Q}_{\gamma} +\textsl{Q}^{\gamma}\delta
\textsl{Q}_{\gamma}\bigg),
\\\nonumber
&=&-\frac{1}{4}\bigg[\textsl{Q}_{\gamma\eta\xi}
\nabla^{\gamma}\delta \mathrm{g}^{\eta\xi}-\textsl{Q}^{\gamma
\eta\xi} \nabla_{\gamma}\delta
\mathrm{g}_{\eta\xi}-2\textsl{Q}_{\xi\gamma\eta} \nabla^{\gamma
}\delta
\mathrm{g}^{\eta\xi}+2\textsl{Q}^{\gamma\eta\xi}\nabla_{\xi}\delta
\mathrm{g}_{\gamma\eta}\\\nonumber
&+&2\tilde{\textsl{Q}}_{\gamma}\nabla^{\gamma}\mathrm{g}^{\xi\eta}\delta
\mathrm{g}_{\xi\eta}+2\tilde{\textsl{Q}}_{\gamma}\mathrm{g}_{\xi\eta}\nabla^{\gamma}\delta
\mathrm{g}^{\xi\eta}-2\textsl{Q}^{\gamma}\nabla^{\vartheta}\delta
\mathrm{g}_{\gamma\vartheta}-\textsl{Q}_{\gamma}\nabla^{\gamma}\mathrm{g}^{\xi\eta}\delta
\mathrm{g}_{\xi\eta}\\
\label{B9}
&-&\textsl{Q}_{\gamma}\mathrm{g}_{\xi\eta}\nabla^{\gamma}\delta
\mathrm{g}^{\xi\eta}-\textsl{Q}^{\gamma}\nabla_{\gamma}\mathrm{g}^{\xi\eta}\delta
\mathrm{g}_{\xi\eta}-\textsl{Q}^{\gamma}\mathrm{g}_{\xi\eta}\nabla_{\gamma}\delta
\mathrm{g}^{\xi\eta}\bigg].
\end{eqnarray}
Here, we use the following equations as
\begin{eqnarray}\label{B10}
\delta \mathrm{g}_{\xi\eta}&=&-\mathrm{g}_{\xi\gamma }\delta
\mathrm{g}^{\gamma\vartheta}\mathrm{g}_{\vartheta\eta},
\\\nonumber
-\textsl{Q}^{\gamma \eta\varepsilon}\nabla_{\gamma}\delta
\mathrm{g}_{\eta\varepsilon}&=&-\textsl{Q}^{\gamma
\eta\varepsilon}\nabla_{\gamma}\big(-\mathrm{g}_{\eta\vartheta}\delta
\mathrm{g}^{\vartheta\vartheta}\mathrm{g}_{\vartheta\varepsilon}\big)
\\\label{B11}
&=&2\textsl{Q}^{\gamma\psi}_{~~\eta}\textsl{Q}_{\gamma\psi\xi}\delta
\mathrm{g}^{\xi\eta}+\textsl{Q}_{\gamma
\eta\varepsilon}\nabla^{\gamma}\mathrm{g}^{\eta\varepsilon},
\\\label{B12}
2\textsl{Q}^{\gamma\eta\varepsilon}\nabla_{\varepsilon}\delta
\mathrm{g}_{\gamma
\eta}&=&-4\textsl{Q}^{~~\psi\varepsilon}_{\xi}\textsl{Q}_{\varepsilon\psi\eta}\delta
\mathrm{g}^{\xi\eta}-2\textsl{Q}_{\eta\varepsilon\gamma}\nabla^{\gamma}\mathrm{g}^{\eta\varepsilon},
\\\nonumber
-2\textsl{Q}^{\varepsilon}\nabla^{\vartheta}\delta
\mathrm{g}_{\varepsilon\vartheta}&=&2\textsl{Q}^{\gamma}\textsl{Q}_{\eta\gamma
\xi}\delta
\mathrm{g}^{\xi\eta}+2\textsl{Q}_{\xi}\tilde{\textsl{Q}}_{\eta}\delta
\mathrm{g}^{\xi\eta}
\\\label{B13}
&+&2\textsl{Q}_{\eta}\mathrm{g}_{\gamma\varepsilon}\nabla^{\gamma}\mathrm{g}^{\eta\varepsilon}.
\end{eqnarray}
Thus, Eq.(\ref{B9}) becomes
\begin{equation}\label{B14}
\delta \textsl{Q}=2\textsl{P}_{\gamma
\eta\varepsilon}\nabla^{\gamma}\delta
\mathrm{g}^{\eta\varepsilon}-\big(\textsl{P}_{\xi\gamma\vartheta}\textsl{Q}^{~~\gamma
\vartheta}_{\eta}
-2\textsl{Q}^{\gamma\vartheta}_{~~\xi}\textsl{P}_{\gamma\vartheta\eta}\big)\delta
\mathrm{g}^{\xi\eta},
\end{equation}
where
\begin{eqnarray}\nonumber
2\textsl{P}_{\gamma\eta\varepsilon}&=&-\frac{1}{4}\bigg[2\textsl{Q}_{\gamma
\eta\varepsilon}-
2\textsl{Q}_{\varepsilon\gamma\eta}-2\textsl{Q}_{\eta\varepsilon\gamma}+2
\textsl{Q}_{\eta}\mathrm{g}_{\gamma \varepsilon}
\\\label{B15}
&+&2(\tilde{\textsl{Q}}_{\gamma}-\textsl{Q}_{\gamma})\mathrm{g}
_{\eta\varepsilon}\bigg],
\\\nonumber
4\big(\textsl{P}_{\xi\gamma\vartheta}\textsl{Q}^{~~\gamma
\vartheta}_{\eta}
-2\textsl{Q}^{\gamma\vartheta}_{~~\xi}\textsl{P}_{\gamma\vartheta\eta}\big)&=&
2\textsl{Q}^{\gamma \vartheta}_{~~\eta}Q _{\gamma\vartheta\xi}-4
\textsl{Q}^{~~\gamma
\vartheta}_{\xi}\textsl{Q}_{\vartheta\gamma\eta}
+2\tilde{\textsl{Q}}^{\gamma}\textsl{Q}_{\gamma\xi\eta}
\\\label{B16}
&+&2\textsl{Q}^{\gamma }\textsl{Q}_{\eta\gamma\xi}
+2\textsl{Q}_{\xi}\tilde{\textsl{Q}}_{\eta}-
\textsl{Q}^{\gamma}\textsl{Q}_{\gamma\xi\eta}.
\end{eqnarray}
\\
\textbf{Data Availability Statement:} The research presented in this
paper did not utilize any data.


\vspace{0.25cm}

\begin{thebibliography}{55}

\bibitem{1} Swaters, R.A. et al.: Astrophys. J. \textbf{531}(2000)107.

\bibitem{2} Sahni, V. and Starobinsky, A.A.: Int. J. Mod. Phys. D \textbf{9}(2000)373;
Carroll, S.M.: Living Rev. Rel. \textbf{4}(2001)1.

\bibitem{3} Nojiri, S.I. and Odintsov, S.D.: Int. J. Geom. Methods Mod. Phys. \textbf{4}(2007)115.

\bibitem{4} Urban, F.R. and Zhitnitsky, A.R.: Phys. Rev. D \textbf{80}(2009)063001.

\bibitem{5} Ohta, N.: Phys. Lett. B \textbf{695}(2011)41.

\bibitem{6} Cai, R.G. et al.: Phys. Rev. D \textbf{84}(2011)123501.

\bibitem{7} Sheykhi, A. and Movahed, M.S.: Gen. Relativ. Gravit. \textbf{44}(2012)449.

\bibitem{8} Feng, C.J. et al.: Mod. Phys. Lett. A \textbf{27}(2012)1250182.

\bibitem{13} Hayashi, K. and Shirafuji, T.: Phys. Rev. D \textbf{19}(1979)3524.

\bibitem{14} Linder, E.V.: Phys. Rev. D \textbf{81}(2010)127301.

\bibitem{17} Jimenez, J.B. et al.: Phys. Rev. D \textbf{98}(2018)044048.

\bibitem{17a} Adeel, M.: et al.: Mod. Phys. Lett. A \textbf{38}(2023)2350152.

\bibitem{17b} Sharif, M.: et al.: Chin. J. Phys. \textbf{91}(2024)66.

\bibitem{17c} Rani, S.: et al.: Int. J. Geom. Methods Mod. Phys.
\textbf{21}(2024)2450033.

\bibitem{17d} Gul, M.Z. et.: Eur. Phys. J. C \textbf{84}(2024)8.

\bibitem{17e} Maurya, S.K. et al.: Phys. Dark Universe \textbf{46}(2024)101619.

\bibitem{17f} Sharif, M.: et al.:  New Astron. \textbf{109}(2024)102211.

\bibitem{17g} Rani, S.: et al.: Phys. Dark Universe \textbf{47}(2025)101754.

\bibitem{17h} Sharif, M.: et al.: Phys. Dark Universe \textbf{47}(2025)101760.

\bibitem{18a} Zhadyranova, A.: J. High Energy Astrophys. \textbf{44}(2024)123.

\bibitem{18b} Gul, M.Z. et al.: Phys. Scr. \textbf{99}(2024)045006.

\bibitem{18c} Koussour, M.: Chin. J. Phys. \textbf{90}(2024)108.

\bibitem{18d} Sharif, M. et al.: Phys. Scr. \textbf{99}(2024)115003.

\bibitem{18dd} Gul, M.Z. et al.: Chin. Phys. C. \textbf{48}(2024)12503.

\bibitem{18e} Koussour, M.: Phys. Dark Universe \textbf{45}(2024)101527.

\bibitem{18f} Gul, M.Z. et al.: Chin.  J. Phys. \textbf{93}(2025)256.

\bibitem{18g} Nan, G. et al.: Phys. Dark Universe
\textbf{46}(2024)101635.

\bibitem{18h} Gul, M.Z. et al.: Eur. Phys. J. C
\textbf{84}(2024)775; ibid 802; ibid 1232.

\bibitem{18i} Sharif, M. et al.: Mod. Phys. Lett. A  \textbf{39}(2024)2450140.

\bibitem{18j} Pradhan, S. et al.: Fortschr. der Phys. \textbf{72}(2024)2400092.

\bibitem{18k} Gul, M.Z. et al.: Gen. Relativ. Gravit. \textbf{56}(2024)45.

\bibitem{19a} Koussour, M. et al.: Phys. Dark Universe \textbf{46}(2024)101577.

\bibitem{19c} Gul, M.Z. et al.: Chin. J. Phys. \textbf{88}(2024)388.

\bibitem{19d} Sharif, M. et al.: Eur. Phys. J. C \textbf{84}(2024)1094.

\bibitem{19e} Myrzakulov, Y. et al.: Phys. Dark Universe \textbf{45}(2024)101545.

\bibitem{19ee} Sharif, M. and Gul, M.Z.:  Ann. Phys. \textbf{465}(2024)169674.

\bibitem{19f} Errehymy, A. et al.: Physics of the Dark Universe \textbf{46}(2024)101555.

\bibitem{19ff} Sharif, M. and Gul, M.Z.:  Phys. Scr. \textbf{99}(2024)065036.

\bibitem{19g} Gul, M.Z. et al.:  Chin. J. Phys. \textbf{89}(2024)1347.

\bibitem{23} Myrzakulov, Y. et al.: Phys. Dark Universe \textbf{46}(2024)101614.

\bibitem{33} Harko, T. et al.: Phys. Rev. D \textbf{98}(2018)084043.

\bibitem{34} Mandal, S. and Sahoo, P.K.: Phys. Lett. B \textbf{823}(2021)136786.

\bibitem{35} Myrzakulov, K.: J. High Energy Astrophys. \textbf{44}(2024)164.

\bibitem{24} Turner, M.S. and White M.: Phys. Rev. D \textbf{56}(1997)R4439.

\bibitem{25} Sahni, V. and Starobinsky, A.: Int. J. Mod. Phys. D \textbf{15}(2006)2105.

\bibitem{27} Chirde, V.R. and Shekh, S.H.: Astron. Astrophys. \textbf{58}(2015)106.

\bibitem{29} Arora, S. et al.: Phys. Dark Universe. \textbf{30}(2020)100664.

\bibitem{30} Solanki, R. et al.: Phys. Dark Universe. \textbf{36}(2022)100996.

\bibitem{31} Mussatayeva, A. et al.: Phys. Dark Universe. \textbf{42}(2023)101276.

\bibitem{38} Ebrahimi, E. and Sheykhi, A.: Phys. Lett. B \textbf{706}(2011)19.

\bibitem{39} Saaidi, K. et al.: Int. J. Mod. Phys. D \textbf{21}(2012)1250057.

\bibitem{41} Jawad, A.: Astrophys. Space Sci. \textbf{356}(2015)119.

\bibitem{42} Fayaz, V. et al.: Eur. Phys. J. Plus \textbf{131}(2016)22.

\bibitem{43} Sharif, M. and Nawazish, I.: Int. J. Mod. Phys. D \textbf{27}(2018)1850091.

\bibitem{52} Saridakis, E.N. et al.: J. Cosmol. Astropart. Phys. \textbf{2018}(2018)012.

\bibitem{53} Zadeh, M.A., Sheykhi, A., Moradpour, H. and Bamba, K.: Eur. Phys. J. C \textbf{78}(2018)11.

\bibitem{54} Ghaffari, S. et al.: Eur. Phys. J. C \textbf{78}(2018)706.

\bibitem{55} Huang, Q. et al.: Class. Quantum Grav. \textbf{36}(2019)175001.

\bibitem{37} Odintsov, S.D., Oikonomou, V.K. and Banerjee, S.: Nucl. Phys. B \textbf{938}(2019)935.

\bibitem{45} Myrzakulov, N. et al.: Front. Astron. Space Sci. \textbf{9}(2022)902552.

\bibitem{49} Sharif, M., Gul, M.Z. and Hashim, I.: Phys. Dark Universe \textbf{46}(2024)101606.

\bibitem{49c} Sharif, M., Gul, M.Z. and Hashim, I.: Chin. J. Phys. \textbf{89}(2024)266.

\bibitem{50} Gul, M.Z., Sharif, M. and Hashim, I.: Phys. Dark Universe \textbf{45}(2024)101537.

\bibitem{63} Gadbail, G.N., Mandal, S. and Sahoo, P.K.: Physics \textbf{4}(2022)1403.

\bibitem{66} Caldwell, R.R. and Linder, E.V.: Phys. Rev. Lett. \textbf{95}(2005)141301.

\bibitem{67} Sahni, V. et al.: J. Exp. Theor. Phys. Lett. \textbf{77}(2003)201.

\bibitem{68} Myrzakulov, N.: Front. Astron. Space Sci. \textbf{9}(2022)902552.

\bibitem{69} Ade, P.A. et al.: Astron. Astrophys. \textbf{594}(2016)13.

\bibitem{70} Sharif, M. and Ajmal, M.: Chin. J. Phys. \textbf{88}(2024)706.

\bibitem{70a} Sharif, M. and Ibrar, I.: Eur. Phys. J. Plus \textbf{139}(2024)17.

\bibitem{70b} Sharif, M. and Zubair, M.: Astrophys. Space Sci. \textbf{353}(2014)699.

\end{thebibliography}
\end{document}